\documentclass[12pt,a4paper]{article}
\usepackage{latexsym,amsmath,amssymb}
\date{}
\numberwithin{equation}{section}

\begin{document}
\title{Definite and Indefinite Inner Product on Superspace\\(Hilbert-Krein Superspace)}
\author{Florin Constantinescu\\ Fachbereich Mathematik \\ Johann Wolfgang Goethe-Universit\"at Frankfurt\\ Robert-Mayer-Strasse 10\\ D 60054
Frankfurt am Main, Germany }
\maketitle

\begin{abstract}
We present natural (invariant) definite and indefinite scalar products
on the $N=1$ superspace which turns out to carry an inherent
Hilbert-Krein structure. We are motivated by supersymmetry in physics but prefer a general mathematical framework.

\end{abstract}

\section{Introduction}

Supersymmetries generalize the notion of a Lie algebra to include
algebraic systems whose defining relations involve commutators as well
as anticommutators. Denoting by $  Q_{\alpha },\bar Q_{\dot \alpha }$
the odd (anticommuting) generators, physical considerations require that
(see \cite {WB}) the operators $ Q_{\alpha }  ,\bar Q_{\dot \alpha
}=(Q_\alpha )^+ $ act in a bona fide Hilbert space $\mathcal{H}$ of
states with positive definite metric. Here $(Q_\alpha)^+$ means the
operator adjoint to $Q_{\alpha } $ in $ \mathcal{H} $.  
From the commutation relations \cite {WB}

\begin{equation}\nonumber
\{Q_{\alpha },\bar Q_{\dot \alpha }\}=2\sigma _{\alpha \dot \alpha }^l P_l
\end{equation}
where $\sigma^l ,l=0,1,2,3$ are the Pauli matrices with $\sigma^0=-1$ as in \cite{WB} and $P_l$ is the momentum, it follows that for any state $\Phi $ in $\mathcal {H}$ we have

\begin{equation}\nonumber
\|Q_{\alpha } \Phi \|^2+\| \bar Q_{\dot \alpha } \Phi \|^2=(\Phi ,\{Q_\alpha ,\bar Q_{\dot \alpha }\}\Phi)=2\sigma_{\alpha \dot \alpha }^l (\Phi ,P_l \Phi )
\end{equation}
Summing over $\alpha =\dot \alpha =1,2 $ and using $ tr \sigma ^0=-2, tr \sigma^l=0 ,l=1,2,3 $ yields for the Minkowski metric $(-1,1,1,1)$

\begin{equation}\nonumber
(\Phi, P_0\Phi )>0
\end{equation}
i.e. in a supersymmetric theory the energy $ H=P_0 $ is always
positive. This positivity argument doesn't require any detailed knowledge
of the Hilbert space $\mathcal {H}$ which is an imperative of any
quantum theory. In this paper we present indefinite but also definite
(invariant) inner products on $N=1$ superspace, i.e. defined on
supersymmetric functions on the $N=1$ superspace and show that the
inherent Hilbert space in supersymmetric theories appears in conjunction
with an indefinite (Krein) scalar product. Roughly speaking each
function on superspace can be decomposed in a chiral, antichiral and a
transversal contribution. It turns out that in order to obtain
positivity of the scalar product the transversal contribution has to be
substracted instead of adding it to the chiral/antichiral part.\\
Despite of the previous
positivity argument leading to the energy positivity which relies on physical arguments, we prefer for
this paper a general mathematical framework and even do not explicitely assume
supersymmetry; in particular we do not assume Lorentz invariance. Comments on physics appear at the end of the paper.
We use the notation and conventions of \cite{WB} with the only difference that from now on $ \sigma ^0, \bar \sigma ^0 $
are the identity instead of minus identity (our notations coincide with
\cite {S}). In particular our Minkowski metric $\eta^{lm}$ is $(-1,+1,+1,+1)$. The Fourier transform $\tilde f  (p)$ of $f(x)$ is defined through
\[ f(x)=\frac {1}{(2\pi )^2}\int e^{ipx}\tilde f (p)dp \]
where $ px=p_l x^l =p_l \eta^{lm }x_m $. \\
We use 
the Weyl spinor formalism in the Van der Waerden notations as in the
references cited above although for our purposes 
4-component spinors would be better suited (see
\cite{Wei}). Working with Weyl spinors we
have to assume for consistency reasons anticommutativity of their components which in our
case are regular (test) functions (or distributions). This will be not the
case at the point we define sesquilinear form (inner products) by integration on superspace connecting to the usual $L^2$-scalar
product on functions. Certainly this is not a serious problem as it is clear to the reader (see also Section 3). 

\section{The supersymmetric functions}

We restrict to the $N=1$ superspace. We write the most general
superspace (test) function $X=X(z)=X(x,\theta ,\bar \theta )$ as in \cite{WB,S}

\begin{gather}\nonumber 
X(z)=X(x,\theta ,\bar \theta )= \\ \nonumber
=f(x)+\theta \varphi (x) +\bar \theta \bar \chi (x) +\theta ^2m(x)+\bar \theta^2n(x)+ \\ 
\theta \sigma^l\bar \theta v_l(x)+\theta^2\bar \theta \bar \lambda(x)+\bar \theta^2\theta \psi (x)+ \theta^2 \bar \theta^2d(x)
\end{gather}
where the coefficients are functions of $x$ in Minkowski space of certain regularity which will be specified below (by the end of the paper we will admit distributions too ). For the time being suppose that the coefficient functions are in the Schwartz space $S$ of infinitely differentiable (test) functions with faster than polynomial decrease at infinity.
For the vector component $v$ we can write equivalently \[\theta \sigma^l\bar \theta v_l=\theta ^{\alpha }\bar \theta ^{\dot \alpha }v_{\alpha \dot \alpha } \]
where \[v_{\alpha \dot \alpha }=\sigma _{\alpha \dot \alpha }^l
v_l,v^l=-\frac{1}{2}\bar \sigma^{l\dot \alpha \alpha }v_{\alpha \dot
  \alpha } \]
which is a consequence of the "second" completeness equation
\[ \sigma_{\alpha \dot \beta }^l \bar \sigma_l^{\dot \gamma \rho }=-2\delta_{\alpha }^{\rho }\delta_{\dot \beta }^{\dot \gamma }. \]
Let us introduce the supersymmetric covariant (and invariant \cite{WB,S})
derivatives $D,\bar D$ with spinorial components $D_{\alpha },D^{\alpha
},\bar D_{\dot \alpha },\bar D^{\dot \alpha }$ given by

\begin{gather}  
D_{\alpha }=\partial_{\alpha } +i\sigma_{\alpha \dot \alpha }^l\bar
\theta^{\dot \alpha }\partial _l \\ 
D^{\alpha }=\epsilon ^{\alpha \beta }D_{\beta }=-\partial ^{\alpha }+i\sigma^{l\alpha }_{\dot \alpha }\bar \theta^{\dot \alpha }\partial _l  \\
\bar D_{\dot \alpha }=-\bar {\partial }_{\dot \alpha } -i\theta ^{\alpha }\sigma _{\alpha \dot \alpha }^l\partial _l \\ 
\bar D^{\dot \alpha }=\epsilon ^{\dot \alpha \dot \beta }\bar D_{\dot
  \beta }=\bar {\partial }^{\dot \alpha }-i\theta ^{\alpha }\sigma
_{\alpha }^{l\dot \alpha }\partial _l   
\end{gather}
We accept on the way notations like
\[\epsilon^{\alpha \beta }\sigma^l _{\beta \dot \alpha }=\sigma^{l\alpha
}_{\dot \alpha } \] etc. but in the end we come back to the canonical index positions $\sigma^l =(\sigma_{\alpha \dot \alpha }^l ),\bar \sigma^l =(\bar \sigma^{l\dot \alpha \alpha })$.\\
Note that $D_{\alpha }$ does not contain the variable $\theta $ and $\bar D ^{\dot \alpha } $ does not contain the variable $\bar \theta $ such that we can write at the operatorial level:

\begin{gather}
D^2=D^{\alpha }D_{\alpha }=-(\partial ^{\alpha }\partial _{\alpha }-2i\partial _{\alpha \dot \alpha }\bar \theta ^{\dot \alpha } \partial ^{\alpha }  +\bar \theta^2 \square ) \\   
\bar D^2=\bar D_{\dot \alpha }\bar D^{\dot \alpha }=-(\bar {\partial } _{\dot \alpha }\bar {\partial }^{\dot \alpha }+2i \theta ^{\alpha }\partial _{\alpha \dot \alpha }\bar \partial ^{ \dot \alpha } +\theta^2\ \square )    
\end{gather}
where 
\[\square =\eta ^{lm}\partial _l \partial _m \]
is the d'alembertian, $\eta $ is the Minkowski metric tensor and
\[ \partial _{\alpha \dot \alpha }=\sigma _{\alpha \dot \alpha }^l \partial _l \]
Here we used the "first" completeness relation for the Pauli matrices $\sigma ,
\bar \sigma $: 

\begin{equation}
Tr(\sigma^l \bar \sigma^m )=\sigma_{\alpha \dot \beta }^l \bar \sigma^{m\dot
  \beta \alpha }=-2\eta^{lm} 
\end{equation}
We make use of the operators \cite{WB,S}

\begin{gather}
c=\bar D^2D^2, a=D^2\bar D^2, T=D^{\alpha }\bar D^2 D_{\alpha }=\bar
D_{\dot \alpha }D^2 \bar D^{\dot \alpha }=-8\square +\frac {1}{2}(c+a) 
\end{gather}
which are used to construct formal projections

\begin{gather}
P_c=\frac {1}{16\square }c,P_a=\frac {1}{16\square }a, P_T=-\frac
{1}{8\square }T 
\end{gather}
on chiral, antichiral and transversal supersymmetric functions. These operators are, at least for the time being, formal
because they contain the d'alembertian in the denominator. Problems with
the d'alembertian in (2.10) in the denominator will be explained later. Chiral, antichiral and transversal funtions are linear subspaces of general supersymmetric functions which are defined by the conditions \cite{WB,S}
\[ \bar D^{\dot \alpha }X=0, \dot \alpha =1,2,;  D^{ \alpha }X=0, \alpha =1,2; D^2 X=\bar D^2 X=0 \]
respectively. 
It can be proven that these relations are formally equivalent to the relations
\[P_cX=X, P_aX=X, P_TX=X \] 
(we mean here that $P_i ,i=c,a,T $ are applicable to $X$ and the relations above hold). \\
We have formally 
\[ P_i^2=P_i, P_iP_j=0,\quad i\ne j;i,j=c,a,T \]
and $P_c +P_a +P_T =1 $. Accordingly each supersymmetric function can be
formally decomposed into a sum of a chiral, antichiral and transversal
contribution (from a rigorous point of view this statement may be wrong
and has to be reconsidered because of the problems with the d'alembertian in the denominator; fortunately we will not run into such difficulties as this will be made clear later in the paper). \\
Let us specify the coefficient functions in (2.1) for the chiral,
antichiral and transversal supersymmetric functions. \\
For the chiral case $X_c$ we have:

\begin{gather} \nonumber 
\bar \chi =\psi =n=0 , v_l=\partial_l (if)=i\partial_l f ,
  \\ \bar \lambda =-\frac {i}{2}\partial_l \varphi \sigma^l
   =\frac{i}{2}\bar \sigma^l \partial _l \varphi , 
  d=\frac{1}{4}\square f 
\end{gather}
Here $ f,\varphi $ and $ m $ are arbitrary functions. For notations and relations see (2.23)-(2.27).\\
For the antichiral $X_a$ case:

\begin{gather} \nonumber
\varphi =\bar \lambda =
m=0,  v_l=\partial_l (-if)=-i\partial_l f , \\ \psi =\frac{i}{2}\sigma^l
\partial_l \bar \chi =-\frac{i}{2}\partial _l \bar \chi \bar \sigma^l , d=\frac{1}{4}\square f 
\end{gather}
Here $f,\bar \chi $ and $ n $ are arbitrary functions. \\
For the transversal case $X_T$ \cite{S}:

\begin{gather} \nonumber
m=n=0, \partial_l v^l =0, \\ 
\bar \lambda =\frac {i}{2}\partial_l \varphi
\sigma^l =-\frac{i}{2}\bar \sigma^l \partial_l \varphi ,\psi
=\frac{i}{2} \partial_l \bar
\chi \bar \sigma^l =-\frac{i}{2}\sigma^l \partial_l \bar \chi  ,
d=-\frac{1}{4}\square f 
\end{gather}
Here $f,\varphi ,\bar \chi $ are arbitrary and $v$ satisfies
$\partial_lv^l=0 $.\\
Later on we will need the $ \theta^2 \bar \theta^2 - $ coefficients $[\bar
X_i X_i ](x_1,x_2)  $ of the
quadratic forms $\bar X_i (x_1,\theta ,\bar \theta )X_i (x_2,\theta ,\bar
\theta ) $ for $i=c,a,T $ where $X_0=X $ is arbitrary
supersymmetric. They coincide with the Grassmann integrals

\begin{equation}
\int d^2\theta_1 d^2 \bar \theta_1  d^2\theta_2 d^2 \bar \theta_2 \bar
X_i (x_1,\theta _1,\bar \theta _1)\delta^2(\theta_1 -\theta_2 )\delta^2
(\bar \theta_1 -\bar \theta_2 ) X_i (x_2,\theta _2,\bar
\theta _2) 
\end{equation}
where $ \delta^2(\theta_1 -\theta_2 )=(\theta_1 -\theta_2 )^2 , \delta^2
(\bar \theta_1 -\bar \theta_2 )=(\bar \theta_1 -\bar \theta_2 )^2 ,d^2\theta =\frac{1}{2}d\theta^1 d\theta^2 ,d^2\bar \theta =-\frac{1}{2}d\bar \theta^{\bar 1} d\bar \theta^{\bar 2}$ and are listed below in the order 
of $i=c,a,T$ :

\begin{gather}\nonumber
[\bar X_c X_c ](x_1,x_2)= 
\bar f(x_1)(\frac{1}{4}\square f(x_2))-\bar \varphi (x_1)(\frac{i}{2}\bar \sigma^l \partial_l \varphi (x_2))+\bar m(x_1)m(x_2)-\\
-\frac{1}{2} \partial^l \bar f(x_1)\partial_l f(x_2)
-(-\frac{i}{2}\partial_l \bar \varphi (x_1)\bar \sigma^l )\varphi (x_2)+(\frac{1}{4} \square \bar f(x_1))f(x_2))
\end{gather}

\begin{gather}\nonumber
[\bar X_a X_a ](x_1,x_2)= 
\bar f(x_1)(\frac{1}{4}\square f(x_2))-\chi (x_1)(\frac{i}{2}\sigma^l
\partial_l  \bar \chi (x_2))+\bar n(x_1) n(x_2)- \\ -
\frac{1}{2} \partial^l \bar f(x_1)\partial_l f(x_2)-
(-\frac{i}{2}\partial_l \chi (x_1)\sigma^l )\bar \chi (x_2)+(\frac{1}{4} \square \bar f(x_1))f(x_2))
\end{gather}

\begin{gather}\nonumber
[\bar X_T X_T ](x_1,x_2)= \nonumber
\bar f(x_1)(-\frac{1}{4}\square f(x_2))-\bar \varphi (x_1)(-\frac{i}{2}\bar
\sigma^l \partial_l \varphi (x_2))-\\ \nonumber
-(\frac{i}{2}\partial_l \bar \varphi
(x_1))\bar \sigma^l \varphi (x_2)-\frac{1}{2}
\bar v^l(x_1)v_l(x_2)- \chi (x_1)(-\frac{i}{2} \sigma^l \partial_l \bar \chi (x_2))-\\
-(\frac{i}{2}\partial_l \chi
(x_1)\sigma^l )\bar \chi (x_2)+(-\frac{1}{4} (\square \bar f(x_1))f(x_2))
\end{gather}
where we have used relations quoted in (2.23)-(2.27). The conjugate $\bar X$ is given in (2.34). \\
As an useful exercise let us put $x_1=x_2 $ in $ [\bar X_i X_i
](x_1,x_2),i=c,a,T $ and compute the integral

\[ \int d^4 x[\bar X_i X_i](x) \]
We want to make clear that this computation is done only for pedagogical 
reasons; we perform it because we will need a similar computation in momentum space (!) at a
later stage in this paper. We integrate by parts and use the faster than
polynomial decrease of the coefficient functions and of their derivatives 
to obtain for the chiral case:

\begin{gather} \nonumber
\int d^4x [\bar X_c X_c](x)= \\ 
=\int d^4x \bar f (x)\square f(x) -\int d^4x \bar
\varphi (x)i\bar \sigma^l \partial_l \varphi (x)+\int d^4x \bar m
(x)m(x) 
\end{gather} 
For the antichiral case:

\begin{gather} \nonumber
\int d^4x [\bar X_a X_a](x)=\\ \int d^4x \bar f (x)\square f(x) -\int d^4x 
\chi (x)i\sigma^l \partial_l \bar \chi (x)+\int d^4x \bar n
(x)n(x)   
\end{gather}
and for the transversal case:

\begin{gather}\nonumber
\int d^4x [\bar X_T X_T ](x) =-\frac{1}{2}\int d^4x \bar f (x)\square f(x) +\\ 
\int d^4x \bar \varphi (x)i\bar \sigma^l \partial_l \varphi (x)+\int d^4x 
\chi (x)i\sigma^l \partial_l \bar \chi (x)-\frac{1}{2} \int d^4x \bar v^l
(x)v_l (x)
\end{gather}
Certainly the best we can expect in our paper is to find a Hilbert space structure on
supersymmetric functions such that the decomposition formally suggested
by $P_c+P_a+P_T=1$ is a direct
orthogonal sum of chiral, antichiral and transversal functions, but this
is definitely not the case as will be clear soon. In this paper we
are going to uncover the exact mathematical structure of this
decomposition in its several variants. This will be done by explicit
computations. We start computing the action of the opertators $D_{\alpha
}, D^{\alpha },\bar D_{\dot \alpha }, \bar D^{\dot \alpha },D^2,\bar
D^2, c,a,T $ on $X$. Usually in physics one doesn't need the results of all
these elementary but long computations in an explicit way and this is
the reason they are not fully recorded in the literature. It turns out
that for our purposes we need at least some of them.\\
For a given $X$ as in (2.1) the expressions  $D_{\beta }X,
D^{\gamma }X,\bar D^{\dot \beta }X, \bar D^{\dot \gamma }X $ are easily
computed but are not given explicitely here because they are in fact not necessary in order to
compute higher covariant derivatives used in this paper (in order to
compute higher derivatives we use (2.6) and (2.7)).\\
We start recording the results for $ D^2,\bar D^2$ applied on $X$:

\begin{gather}\nonumber
\bar D^2 X=-4n+\theta (-4\psi -2i\sigma^l \partial_l \bar \chi )
+\theta^2 (-4d -2i\partial _l v^l-\square f)+\theta \sigma^l \bar \theta
(-4i\partial _l n)+\\ \nonumber
 +\theta^2 \bar \theta (-2i\bar \sigma^l \partial _l \psi  -\square \bar \chi ) +\theta^2 \bar \theta^2 (-\square n) \\ \nonumber
D^2 X=-4m+\bar \theta (-4\bar \lambda -2i\bar \sigma^l \partial_l \varphi )+\bar \theta ^2(-4d +2i\partial _l v^l-\square f)+\theta \sigma^l \bar \theta (4i\partial _l m)+\\ \nonumber
+\bar \theta^2 \theta (-2i\sigma^l\partial_l \bar \lambda -\square \varphi )+\theta^2 \bar \theta^2 (-\square m) 
\end{gather}
or in a more suggestive way taking into account the
chirality/antichirality of $\bar D^2 X,D^2 X $ (see (2.11), (2.12)):

\begin{gather}\nonumber
\bar D^2 X=-4n+\theta (-4\psi -2i\sigma^l \partial_l \bar \chi ) +\theta^2 (-4d -2i\partial _l v^l-\square f)+\theta \sigma^l \bar \theta (-4i\partial _l n)+\\ 
 +\theta^2 \bar \theta (\frac{1}{2}i\bar \sigma^l \partial_l )(-4\psi -2i\sigma^l \partial_l \bar \chi ) +\theta^2 \bar \theta^2 (-\square n) \\ \nonumber
D^2 X=-4m+\bar \theta (-4\bar \lambda -2i\bar \sigma^l \partial_l \varphi )+\bar \theta ^2(-4d +2i\partial _l v^l-\square f)+\theta \sigma^l \bar \theta (4i\partial _l m)+\\ 
+\bar \theta^2 \theta (\frac{1}{2}i\sigma ^l \partial_l )(-4\bar \lambda -2i\bar \sigma^l \partial_l \varphi ) +\theta^2 \bar \theta^2 (-\square m) 
\end{gather}
We have used the following notations and relations (see for
instance the standard references mentioned above): 

\begin{gather}
 (\psi \sigma^l )_{\dot \beta }
=\psi^{\alpha }\sigma _{\alpha \dot \beta }^l, (\sigma ^l \bar \chi
)_{\beta }=\sigma _{\beta \dot \rho }^l \bar \chi ^{\dot \rho },
(\bar \chi \bar \sigma^l)^{\alpha } =\bar \chi_{\dot \rho } \bar \sigma 
^{l\dot \rho \alpha },  (\bar
\sigma^l \psi )^{\dot \alpha }
=\bar \sigma^{l\dot \alpha \beta } \psi_\beta 
\end{gather}
with $ (\sigma^l \bar
\chi )^{ \alpha }=-(\bar \chi \bar \sigma^l )^{\alpha } $
etc. as well as 

\begin{gather}
\psi \sigma^l \bar 
\chi =\psi^{\alpha }\sigma _{\alpha \dot \beta }^l \bar \chi^{\dot \beta
  }=-\bar \chi \bar \sigma^l \psi =-\bar \chi _{\dot \alpha }\bar
  \sigma^{l\dot \alpha \beta }\psi _{\beta }  
\end{gather}
where $\bar
  \sigma^l_{\dot \alpha \beta }=\sigma^l_{\beta \dot \alpha } $.\\
As far as the complex conjugation is concerned we have:
 
\begin{gather}
(\psi \sigma
^l )^* _{\dot \alpha } =(\sigma^l \bar \psi )_\alpha ,(\bar \chi \bar
\sigma^l )^{\alpha *}=(\bar \sigma \chi )^{\dot \alpha }, (\psi \sigma^l 
\bar \chi )^*=\chi \sigma^l \bar \psi 
\end{gather}
where $*$ is the complex
conjugation defined such that 

\begin{gather}
(\psi^\alpha )^*=\bar \psi^{\dot \alpha }\\
(\psi_\alpha )^*=\bar \psi_{\dot \alpha }  
\end{gather}
The unusual
properties of the Grassmann derivative were taken into account; in
particular $ \partial_{\alpha }^* =-\bar \partial _{\dot \alpha }$ etc..\\  
As expected $ \bar D^2 X $ and $D^2 X$ are chiral and antichiral functions respectively.
We continue with $c=\bar D^2 D^2, a=D^2 \bar D^2 $:

\begin{gather}\nonumber
cX=\bar D^2 D^2 X=16d-8i\partial _l v^l+4\square f+\theta (8\square \varphi +16i\sigma^l \partial _l \bar \lambda )+ \\ \nonumber
+\theta^2 (16\square m)+\theta \sigma^l \bar \theta (16i\partial_l d+8\partial_l \partial_m v^m +4i\partial_l \square f)+ \\ 
\theta^2 \bar \theta (8\square \bar \lambda +4i\bar \sigma^l \partial_l \square \varphi )+\theta^2 \bar \theta^2 (4\square d -2i\partial_l \square v^l + \square^2 f)
\end{gather}

\begin{gather}\nonumber
aX=D^2 \bar D^2 X=16d+8i\partial _l v^l+4\square f+\bar \theta (8\square
\bar \chi +16i\bar \sigma^l \partial _l \psi )+ \\ \nonumber
+\bar \theta^2 (16\square n)+\theta \sigma^l \bar \theta (-16i\partial_l d+8\partial_l \partial_m v^m -4i\partial_l \square f)+ \\ 
\bar\theta^2 \theta (8\square \psi +4i \sigma^l \partial_l \square \bar \chi )+\theta^2 \bar \theta^2 (4\square d +2i\partial_l \square v^l + \square^2 f)
\end{gather}
and finally obtain $T=-8\square +\frac{1}{2}(c+a)$ applied on $X$ as follows:

\begin{gather}\nonumber
TX= 16d-4\square f +\theta (-4\square \varphi +8i\sigma^l \partial _l \bar \lambda )+\bar \theta (-4\square \bar \chi  
+8i\bar \sigma^l \partial_l \psi )+\\ \nonumber
+\theta \sigma^l \bar \theta (8\partial_l \partial^m v_m -8\square v_l )+\theta^2 \bar \theta (-4\square \bar \lambda +2i\bar \sigma^l \partial_l \square \varphi )+\\ 
+\bar \theta^2 \theta  (-4\square \psi +2i\sigma^l \partial_l \square \bar \chi ) +\theta^2 \bar \theta^2 (-4\square d +\square^2 f)
\end{gather}
or
\begin{gather}\nonumber
TX= 16d-4\square f +\theta (-4\square \varphi +8i\sigma^l \partial _l \bar \lambda )+\bar \theta (-4\square \bar \chi
+8i\bar \sigma^l \partial_l \psi )+\\ \nonumber
+\theta \sigma^l \bar \theta (8\partial_l \partial^m v_m -8\square v_l
)+\theta^2 \bar \theta (-\frac{i}{2}\bar \sigma^l \partial_l )(-4\square \varphi +8i\sigma^l \partial _l \bar \lambda ) +\\ 
+\bar \theta^2 \theta (-\frac{i}{2}\sigma^l \partial_l ) (-4\square \bar 
\chi+8i\bar \sigma^l \partial _l \psi )+\theta^2 \bar \theta^2 (-4\square d +\square^2 f)
\end{gather}
Here we have used the relations

\begin{equation}
(\sigma \partial )(\bar \sigma \partial )=(\bar \sigma \partial )(\sigma
\partial )=-\square 1_{2\times 2}
\end{equation}
where we briefly write 

\begin{equation}
\sigma \partial =\sigma^l \partial_l , \bar \sigma \partial
=\bar \sigma^l \partial_l 
\end{equation} 
The relation (2.32) as well as the relation (2.8) follows from

\[\sigma^l \bar \sigma^m +\sigma^m \bar \sigma^l =-2\eta^{lm} 1_{2\times 
  2} \] 
where $1_{2\times 2}$ is the unit $2\times 2$ matrix.  
Written in the spinor notation it reads

\[\sigma_{\alpha \dot \alpha }^l \bar \sigma^{m\dot \alpha \beta }
+\sigma_{\alpha \dot \alpha }^m \bar
\sigma^{l\dot \alpha \beta } =-2\eta^ {lm} \delta_{\alpha }^{\beta } \]
As expected $\bar D^2 D^2 X$ is chiral, $D^2 \bar D^2 X$ is antichiral
and $TX$ is transversal. The transversality (2.13) of $TX$
was put in evidence in (2.31).\\
In order to construct inner products in integral form we also need the
conjugates $\bar X, \overline {\bar D^2 X}, \overline {D^2 X},  $ etc. of $X, \bar
D^2 X, D^2X  $ etc. where the conjugation includes besides the
usual complex conjugation the Grassman conjugation too. We have

\begin{gather} \nonumber
\bar X=\bar X(x,\theta ,\bar \theta )=\\ \nonumber
=\bar f(x)+\theta \chi (x) +\bar \theta \bar \varphi (x) +\theta ^2 \bar n(x)+\bar \theta^2 \bar m(x)+ \\ 
+\theta \sigma^l \bar \theta \bar v_l(x)+\theta^2 \bar \theta \bar \psi(x)+\bar \theta^2\theta \lambda (x)+ \theta^2 \bar \theta^2 \bar d(x)
\end{gather}
where $\bar f,\chi ,\bar \varphi, $ etc. are the complex conjugate
functions to $f,\bar \chi ,\varphi, $ etc..
Note that if $X$ is chiral then $\bar X$ is antichiral and viceversa. If $X$ is transversal than $\bar X $ is transversal. Although not absolutely necessary we record  here other
expressions too which
can be used to give alternative proofs of results to follow by making
use of partial integration in superspace.
They are (use $ (\chi
\sigma^l \bar \psi )^* =\psi \sigma^l \bar \chi $ where $ * $ is the
complex conjugation which could have been written as bar too):

\begin{gather}\nonumber
\overline {\bar D^2 X}=D^2\bar X=-4\bar n+\bar \theta (-4\bar \psi -2i\bar
\sigma^l \partial_l 
\chi ) +\bar \theta^2 (-4\bar d +2i\partial _l\bar  v^l-\square
\bar f)+\\
+\theta \sigma^l \bar \theta (4i\partial _l \bar n) 
 +\bar \theta^2  \theta (-2i \sigma^l \partial_l \bar \psi -\square \chi )
 +\theta^2 \bar \theta^2 (-\square \bar n) \\ \nonumber
\overline{D^2 X}=\bar D^2\bar X=-4\bar m+ \theta (-4\lambda -2i\sigma^l \partial_l \bar
\varphi ) + \theta ^2(-4\bar d -2i\partial _l \bar v^l -\square \bar 
f)+\\
+\theta \sigma^l \bar \theta (-4i\partial _l \bar m) 
+ \theta^2 \bar \theta (-2i\bar \sigma^l \partial_l \lambda -\square \bar
\varphi )+\theta^2 \bar \theta^2 (-\square \bar m) 
\end{gather}
or in a more suggestive way as chiral and antichiral function respectively

\begin{gather}\nonumber
\overline {\bar D^2 X}=-4\bar n+\bar \theta (-4\bar \psi -2i\bar
\sigma^l \partial_l  \chi ) +\bar \theta^2 (-4\bar d +2i\partial _l \bar 
v^l-\square \bar f)+\theta \sigma^l \bar \theta (4i\partial _l \bar n)+\\ 
 +\theta^2 \bar \theta (\frac{i}{2} \sigma^l \partial_l )(-4\bar \psi
 -2i\bar \sigma^l \partial_l \chi ) +\theta^2 \bar \theta^2 (-\square \bar n) \\ \nonumber
\overline {D^2 X}=-4\bar m+ \theta (-4 \lambda -2i\sigma^l
\partial_l \bar \varphi )+ \theta ^2(-4\bar d -2i\partial _l \bar
v^l-\square \bar f)+\theta \sigma^l \bar \theta (4i\partial _l \bar m)+\\ 
+ \theta^2 \bar \theta (\frac{i}{2}\bar \sigma ^l \partial_l )(-4
\lambda -2i\sigma^l \partial_l \bar \varphi ) +\theta^2 \bar
\theta^2 (-\square \bar m) 
\end{gather}
Further
\begin{gather}\nonumber
\overline {cX}=\overline {\bar D^2 D^2 X}=16\bar d+8i\partial _l \bar v^l+4\square \bar
f+\bar \theta (8\square \bar \varphi +16i\bar \sigma^l \partial_l \lambda  )+ \\ \nonumber
+\bar \theta^2 (16\square \bar m)+\theta \sigma^l \bar \theta
(-16i\partial_l \bar d+8\partial_l \partial^m \bar v_m -4i\partial_l
\square \bar f)+ \\ 
\bar \theta^2 \theta (8\square \lambda +4i\sigma^l \partial_l \square \bar
\varphi )+\theta^2 \bar \theta^2 (4\square \bar d
+2i\partial_l \square \bar v^l + \square^2 \bar f)
\end{gather}

\begin{gather}\nonumber
\overline {aX}=\overline {D^2 \bar D^2 X}=16\bar d-8i\partial _l \bar v^l+4\square \bar 
f+ \theta (8\square \chi +16i\sigma^l \partial_l \bar \psi  )+ \\ \nonumber
+ \theta^2 (16\square \bar n)+\theta \sigma^l \bar \theta (16i\partial_l
\bar d+8\partial_l \partial^l \bar v_m +4i\partial_l \square \bar f)+ \\ 
\theta^2 \bar \theta (8\square \bar \psi +4i\bar \sigma^l \partial_l \square \chi )+\theta^2 \bar \theta^2 (4\square \bar d -2i\partial_l \square \bar
v^l + \square^2 \bar f)
\end{gather}
and finally

\begin{gather}\nonumber
\overline {TX}=\bar T\bar X=T\bar X = 16\bar d-4\square \bar f + \theta (-4\square  \chi
+8i\sigma^l \partial_l \bar \psi )+\bar \theta (-4\square \bar \varphi +  \\ \nonumber
+8i\bar \sigma^l \partial_l \lambda )
+\theta \sigma^l \bar \theta (8\partial_l \partial^m \bar v_m -8\square
\bar v^l )+\theta^2 \bar \theta (-4\square \bar \psi +2i\bar \sigma^l \partial_l
\square \chi )+\\ 
+\bar \theta^2 \theta  (-4\square \lambda +2i\sigma^l \partial_l \square \bar
\varphi ) +\theta^2 \bar \theta^2 (-4\square \bar d
+\square^2 \bar f)
\end{gather}
or

\begin{gather}\nonumber
\overline {TX}= 16\bar d-4\square \bar f + \theta (-4\square  \chi
+8i\sigma^l \partial_l \bar \psi \bar )+\bar \theta (-4\square \bar \varphi  + \\ \nonumber
+8i\bar \sigma^l \partial_l \lambda )
+\theta \sigma^l \bar \theta (8\partial_l \partial^m \bar v_m -8\square
\bar v_l )+\theta^2 \bar \theta (-\frac{i}{2}\bar \sigma^l \partial
_l)(-4\square \chi +8i \sigma^l \partial_l \bar \psi )+\\ 
+\bar \theta^2 \theta (-\frac{i}{2} \sigma^l \partial _l) (-4\square
\bar \varphi +8i\bar \sigma^l \partial_l \lambda ) +\theta^2 \bar \theta^2 (-4\square \bar d
+\square^2 \bar f)
\end{gather}
We start to look for (invariant) supersymmetric kernel functions $K(z_1,z_2)=K(x_1,\theta_1 ,\bar \theta_1 ; x_2,\theta_2 ,\bar \theta_2 )$ which formally induce inner products on supersymmetric functions by
\begin{equation}
(X_1,X_2)=\int d^8z_1d^8z_2\bar X_1(z_1)K(z_1,z_2)X_2(z_2)=\int \bar X_1KX_2
\end{equation}
where the bar on the r.h.s means conjugation (including Grassmann),
$z_i=(x_i,\theta_i,\bar \theta_i )$ and $d^8z=d^4xd^2\theta d^2\bar
\theta $. In the r.h.s of the last equality we have used a sloppy but
concice notation of the integral under study. The simplest choice for
$K$ would be the identity kernel
$K(z_1,z_2)=k(z_1-z_2)=\delta^2 (\theta_1 -\theta_2)\delta^2 (\bar
\theta_1 -\bar \theta_2 )\delta ^4 (x_1-x_2) $ but it turns out that
this choise is not sound. We settle down soon for more appropriate
choices. Formally we have if $\bar K=K$:
\[ \overline {(X_1,X_2)}=(\bar X_2,\bar X_1) \]
where the bars include Grassmann conjugation. 
The action of $K$ on $X$ is defined formally as
\[ Y_K (z_1)=(KX)(z_1)=\int d^8z_2 K(z_1,z_2)X(z_2)\]
Note that the general dependence of $K$ on $z_1 ,z_2 $ we admit is not necessarily through the difference $z_1 -z_2 $. We assume that the coefficient functions of the
supersymmetric functions involved belong to the Schwartz function space
$S$ of infinitely differentiable rapidly decreasing functions. \\
Now we are starting to induce positivity of the inner product by a proper choice of the kernel $K$. By positivity in this section we mean non-negativity. The first candidate is
\begin{equation}
K(z_1 ,z_2)=K_0(z_1 -z_2)=\delta^2 (\theta_1 -\theta_2)\delta^2 (\bar \theta_1 -\bar \theta_2 )D^+(x_1-x_2)
\end{equation}
where $\delta^2 (\theta_1-\theta_2)=(\theta_1-\theta_2)^2, \delta^2 (\bar
\theta_1-\bar \theta_2)=(\bar \theta_1-\bar \theta_2)^2 $ are the
supersymmetric $\delta $-functions and $D^+ (x)$ is the Fourier transform
of a positive measure $d\rho(p)$ supported in the backward light cone $\bar V^- $
(not necessarily Lorentz invariant):
\begin{equation}
D^+(x)=\frac{1}{(2\pi )^2}\int e^{ipx}d\rho (p)
\end{equation}
which is of polynomial growth i.e. there is an integer $n$ such that 

\begin{equation}
\int \frac{d\rho (p)}{(1+|p|^2)^n}< \infty    
\end{equation}
where $ |p|=\sqrt {p_0^2
  +p_1^2 +p_2^2 +p_3^2} $ . 
Here $px$ is the "most positive" Minkowski scalar product. Usually (for
instance in quantum field theory) the Minkowski scalar product is "most
negative" and as a consequence the measure $d\rho (p)$ is concentrated in 
the forward light cone $ \bar V^+ $. For the time being we do not necessarily assume Lorentz invariance of the measure. The
special kernel $K_0 $ depends only on the difference $z_1 -z_2 $. In order to understand the idea behind this choice note first that for $f$ and $g$ functions of $x$ in $S$ the integral

\begin{equation}
(f,g)=\int d^4xd^4y \bar f(x)D^+(x-y)g(y)
\end{equation}
where $D^+(x)$ is given by (2.45) induces a positive definite scalar
product (certainly in order to exclude zero-vectors we have to require
the support of $f$ and $g$ in momentum space in $\bar V^- $ to be concentrated 
on the support of $d\rho (p) $ which is
equivalent with factoring out zero-vectors and completion in
(2.47)). Indeed the right hand side of (2.47) equals in momentum space
$\int \bar {{\tilde f}}(p)\tilde g(p)d\rho (p) $ where 
$\tilde f $ is the Fourier transform of $f$ given by $f(x)=\frac{1}{(2\pi
  )^2}\int e^{ipx}\tilde f(p)dp $. Note further that 
positivity is preserved if we multiply the measure $d\rho (p)$ by $-p^2$
or for the case of two-spinor functions $f$ and $g$ by $\sigma p $ or
$\bar \sigma p$. In configuration space it means that we can accomodate
the operators $\square $ and $-i\sigma \partial ,-i\bar \sigma \partial $ in
the kernel of the integral without spelling out the positivity (we 
have as usually $\frac{1}{i}\partial =p$).\\
It is clear that in spite of the positivity properties induced by the
kernel $D^+$ the scalar product in (2.43) with kernel (2.44) cannot be
positive definite in superspace for general coefficient functions (for
$X_1=X_2 =X$) because the coefficient functions are mixed up in the process
of Grassmann integration in an uncontrolled way. Fortunately there are other kernels deduced from $K_0$ which do the job. In order to keep the technicalities aside for the moment let us assume that the measure $d\rho (p) $, besides being of polynomial growth, satisfies the stronger condition
\begin{equation}
|\int \frac {1}{p^2} \frac{d\rho (p)}{(1+|p|^2)^n}| < \infty 
\end{equation}
with the integer $n$ appearing in (2.46). \\
Certainly the condition above is relatively strong; it allows measures
like $d\rho (p)=\theta (-p_0)\delta (p^2+m^2)dp$ with $m>0$ but excludes
the case $m=0$ (in physics the massive and massless case
respectively). The case $m=0$ will be studied at the end of this section. \\
We arrived at the level of explaning our message. For this we introduce besides $K_0 (z_1 -z_2)$ three other kernels as follows
\begin{gather}
K_c(z_1,z_2)=P_cK_0 (z_1-z_2)\\
K_a(z_1,z_2)=P_aK_0 (z_1-z_2)\\
K_T(z_1,z_2)=-P_TK_0 (z_1-z_2)
\end{gather}
with actions 
\[ Y_i (z_1)=(K_iX)(z_1)=\int d^8z_2 K_i (z_1,z_2)X(z_2)\]
In (2.49)-(2.51) the operators $P_i $ are understood to act on the first
variable $z_1 $ (see also (2.52)-(2.57) to follow).
The condition (2.48) makes the formal definition $Y=\int KX $ (with $K$
replaced by one of the derived kernels $K_i$ ,$i=c,a,T $ as written
above) safe from a rigorous point of view because it takes care of the
d'alembertian in the denominators introduced by the formal projections
$P_i$, $i=c,a,T$. We will remove this condition soon by slighty
restricting the set of supersymmetric (test) functions but let us keep
it for the time being. Note that the projections destroy the translation 
invariance in the Grassmann variables but not in the space
coordinates. Because $P_i,i=c,a,T $ contain Grassmann variables and
derivatives thereof we have to specify on which variables they act in
$K_0(z_1-z_2)$. By convention let us define by $D_1^2K_0(z_1-z_2), \bar
D_1^2K_0(z_1-z_2),T_1 K_0(z_1-z_2) $ the
action of the operators $D^2 $ and $\bar D^2,T $ on $K_0(z_1-z_2)$ on
the first variable respectively and by  $ D_2^2 K_0(z_1-z_2), \bar D_2^2 K_0(z_1-z_2),  T_2 K_0(z_1-z_2)$ the action of these
operators on the second variable. If indices are not specified, we
understand the action on the first variable.   \\
It can be proven (for similar computations see for instance \cite{S}) that

\begin{gather} 
D_1^2 K_0(z_1-z_2)=D_2^2 K_0(z_1-z_2) \\
\bar D_1^2 K_0(z_1-z_2)=\bar D_2^2 K_0(z_1-z_2)\\ 
D_1^2 \bar D_1^2 K_0(z_1-z_2)=\bar D_2^2 D_2^2 K_0(z_1-z_2)\\
\bar D_1^2 D_1^2 K_0(z_1-z_2)=D_2^2 D_2^2 K_0(z_1-z_2)\\ 
T_1K_0 (z_1-z_2)=T_2K_0 (z_1-z_2) \\
\bar T_1K_0 (z_1-z_2)=\bar T_2K_0 (z_1-z_2)
\end{gather}
where in fact the relations (2.56), (2.57) coincide because $\bar T =T$. We have used 

\[ [D_1^2 ,D_2^2 ]=0,\quad [\bar D_1^2  ,\bar D_2^2 ]=0 \]
Note the minus sign in front of $P_T$ in (2.51) which will be of utmost
importance for us. Because of it the kernels $K_i$, $i=c,a,T $ do not
sum up to $K$. This is at the heart of the matter being at the same time
not too much embarassing.  We will prove by direct computation that the
kernels $K_i,i=c,a,T$ produce, each for itself, a positive definite
scalar product in the space of supersymmetric functions (at this stage
we prove only nonnegativity; the problem of zero vectors is pushed to
Section 3). Whereas this
assertion is to be expected for $K_i$ for $i=c,a$, the minus sign in
$K_T$ comes as a surprise. It will be the reason for the natural Krein (more precisely Hilbert-Krein) structure of the $N=1$ supersymmetry which we are going to uncover
(first under the restrictive condition (2.48) on the measure). Denoting
by $(.,.)_i, i=0,c,a,T$ the inner products induced by the kernels $K_i,
i=0,c,a,T$:

\begin{equation}
(X_1 ,X_2 )_i=\int \bar X_1 K_i X_2
\end{equation} 
we could compute them by brute force using the expressions
(2.28)-(2.30) but it is not easy to get the positive definiteness of these inner products in the cases $i=c,a,T$. Alternatively we will proceed as follows. Let us start with the
cases $i=c,a$. We use (2.52)-(2.57) and integrate by
parts in superspace (see for instance \cite{S}). This gives (in the sloppy integral
notation) by partial integration in superspace 

\begin{gather}
(X_1 ,X_2 )_c =\int \bar X_1  K_c X_2 =\int \bar X_1 P_c K_0 X_2 =(D_1^2X_1,\frac{1}{16\square
}D_2^2X_2)_0 \\  (X_1 ,X_2 )_a =\int \bar X_1
 K_a X_2=\int \bar X_1 P_a K_0 X_2 =(\bar D_1^2X_1,\frac{1}{16\square }\bar D_2^2X_2)_0 
\end{gather}
where we have also used $ \overline {D^2X}=\bar D^2 \bar X $ etc. The last
equality follows from obvious ones supplemented by
$\overline {i\sigma^l \partial_l \bar \varphi }=(i\sigma^l \partial_l \bar
\varphi )^*=-i\partial_l \varphi \sigma^l =i\bar \sigma^l \partial_l
\varphi $, etc.. In (2.59) $\bar D^2 $ from $P_c $ was moved to $\bar
X_1 $, the remaining $D^2 $ (acting on $K_0 $ on the first variable)
was transfered by (2.52) to the second variable on $K_0 $, and then moved on 
$X_2 $ such that finally we get the last expression. The same procedure was applied
for (2.60). The d'alembertian in
the denominator can be absorbed in Fourier space in the measure $d\rho
(p)$ which is supposed to satisfy condition (2.48).
Using the $\delta $-function property in the Grassmann
variables in $K_0 $ we see that for instance in the antichiral case we get for $X_1=X_2=X $

\begin{gather} \nonumber
(X ,X )_a =\int\bar X  K_a X = \\ 
=\int d^4x_1 d^4x_2[\overline {(\bar D^2 X )}
(\bar D^2 X)](x_1,x_2)\frac{1}{16\square}D^+(x_1-x_2)
\end{gather}
where $[.]$, as before, gives the coefficient of the highest power in the
Grassmann variables. \\
Note that $\bar D^2 X$ is chiral such that for $ [\overline {(\bar D^2 X )}
(\bar D^2 X)](x_1,x_2) $ we can apply (2.15). We integrate by parts in the usual coordinates using the 
faster than polynomial decrease of the involved functions and their
derivatives and obtain in momentum space

\begin{gather}\nonumber
\int\bar X  K_a X = \\
=\int[\overline {\tilde f_c }(p)\tilde f_c (p)+\overline {\tilde \varphi_c}(p)(\sigma p)\tilde \varphi _c +\overline {\tilde m_c }(p)\tilde m_c (p))] (\frac{1}{-p^2})d\rho (p)
\end{gather}
where $f_c ,\varphi_c ,m_c $ are the coefficients of the chiral $\bar
D^2 X$ given by (2.21). We have used the translation invariance of $D^+
(x)$ which enables us to read up the result in momentum space from the
computation conducting to (2.18) which was performed in coordinate
space (this is an unusual way to keep track of the $\delta $-function in
momentum space generated by translation invariance which quickly gives the result). \\
From (2.62)  we obtain by inspection the positivity of $\int\bar X K_a
X=(X,P_a X)_0
$. We use the positivity of $-p^2 ,\sigma p $ and $ \bar \sigma p $. The
same argument works for the chiral integral $\int\bar X K_c X=(X,P_c
X)_0
$.\\
Now we go over to the transversal integral $\int\bar X_1 K_T X_2 $. Here we
cannot split the kernel in a useful way as we did in the chiral and
antichiral cases but the following similar procedure can be applied.\\ We
write using $ P_T^2 =P_T $, the relation (2.56) and integration by parts in superspace

\begin{gather}\nonumber
-(X_1 ,X_2 )_T =-\int\bar X_1 K_T X_2 =\int \bar X_1 P_T K_0 X_2 =\int \bar
X_1 P_T^2 K_0 X_2 = \\
=(P_TX_1 ,P_T X_2 )_0=\frac{1}{64}(\frac{1}{\square }TX_1 , \frac{1}{\square }TX_2)_0
\end{gather}
Here, as in the antichiral case above, one of $ P_T $ in $P_T^2 $ acting 
on the first variable was
moved to $\bar X_1 $ and the second one was pushed through $K_0$ (modulo 
changing the variable) to $X_2
$. In (2.63) we take $X_1=X_2=X$, integrate the $\theta ,\bar \theta $-variables and use for $[(\overline {TX})(TX))(x_1,x_2)$ the expression (2.17).
We can use now (2.20) by analogy in momentum space as above too. Note that by integration by parts we 
have enough derivatives in the numerator in order to cancel one of the
two inverse d'alembertians in (2.63). By (2.48) the second d'alembertian is under control and the computation is
safe. We propose to the reader to go this way in order to explicitely
convince himself that the integral $-\int \bar X_1 K_T X_2 =\int \bar X_1 P_T 
K_0 X_2 $ (in contradistinction to the
chiral/antichiral case) is negative for $X_1=X_2 $! A hint is
necessary. Indeed the only contribution which has to be looked up beyond 
the chiral/antichiral case  is the vector contribution stemming from
$v$-coefficients of the transversal supersymmetric function and this produces a negative contribution. In fact the negativity of the transversal contribution rests on the following property in momentum space. Let $v(p)=(v_l (p))$ be a
vector function (not necessary real) such that $p_l v^l (p)=0 $. It means
that the vector with components $v_l (p)$ is orthogonal (in the euclidean
meaning) to the (real) vector $p_l $. But the momentum vector $p$ is
confined to the light cone (it must be in the support of $d\rho (p)$)
such that the vector fuction $v(p)$ must satisfy $\bar v^l (p)v_l
(p)\geq 0$. Moreover if $d\rho $ intersects the light cone $p^2 =0$ the
equality may be realized. We repeat here an old argument which was recognized in
the frame of the rigorous version of the Gupta-Bleuler quantization in
physics \cite{WG,StroW}. But there is a new aspect: whereas in \cite{WG,StroW} the 
free divergence condition $p_l v(p)=0$ was introduced ad hoc in order
to force the Gupta-Bleuler definite metric, it comes in for free here as a
consequence of supersymmetry. \\
The last part of this section is dedicated to the more delicate question
of abolishing the unpleasant restrictive condition (2.48) such that we can include in our analysis, from a physical point of view, the interesting "massless" case. From the consideration above it is clear that this is generally not possible. More precisely, if we want to retain the interpretation of supersymmetric quantum fields as operator-valued (super)distributions
as this is the case for the usual quantum fields \cite{StreW} (an interpretation which we subscribe to) we are forced to restrict the set of allowed test functions such that the d'alembertian in the denominator is annihilated. Restricting the set of test funtions in quantum field theory is not a problem and is not at all new; it appeared even long time ago in the rigorous discussion of the Gupta-Bleuler quantization \cite{WG,StroW}.\\
Suppose that the coefficient functions in (2.1) satisfy the following restrictive conditions:

\begin{gather}
d(x)=\square D(x) \\
\bar \lambda (x)=i\bar \sigma^l \partial_l \Lambda (x) \\
\psi (x) =i\sigma^l \partial_l \bar \Psi (x)\\
v(x)=grad\rho (x)+\omega (x),div\omega (x)=0 
\end{gather}
where $D(x),\Lambda (x),\Psi (x),\rho (x),\omega (x) $ 
are arbitrary functions (in $S$). In the last equation $grad\rho
=(\partial_l \rho ),div\omega =\partial_l \omega^l $.\\
The functions $\rho (x),\omega (x) $ can be constructed as follows: let
$\rho $ be a solution of $\square \rho =div v $ and let $\omega =v-grad
\rho $. Then $v=grad \rho +\omega $ with $div \omega =div v -\square
\rho =0$.\\
We claim that under these conditions the results above concerning the positivity in the chiral/antichiral sectors and negativity in the transversal sector remain valied without the restrictive condition (2.48) on the measure $d\rho $. The conditions (2.64) to (2.67) produce the missing d'alembertian in

\begin{equation}
\int \bar X_1 K_i X_2 ,i=c,a,T 
\end{equation}
such that the condition (2.48) becomes superfluous. Indeed let us consider for example the chiral case (with the antichiral kernel $K_a )$. From (2.21) we see that the following expressions appear in the integral (2.62):

\begin{gather}\nonumber
(-4\bar n)\square (-4n) \\ \nonumber
(-4\bar \psi -2i\bar
\sigma^l \partial_l  \chi )(i\bar \sigma^n \partial_n )(-4\psi -2i\sigma^m \partial_m \bar \chi )\\  \nonumber 
(-4\bar d +2i\partial _l\bar  v^l-\square
\bar f)(-4d -2i\partial _m v^m-\square f)
\end{gather}
It is clear that under the conditions (2.64) to (2.67) the missing
d'alembertian in the integral (2.60) can be factorized such that the
condition (2.48) on the measure $d\rho $ is no longer needed. The result remains positive. Similar
arguments work for the chiral and transversal case. In the
transversal case the interference between $\rho $ and $\omega $ in $\bar 
v^l v_l $ disappears (because $div\omega =0$) and one can use
(besides the positivity of the d'alembertian) again the Gupta-Bleuler
argument with $div\omega =0$. \\
The problem of possible zero-vectors for the non-negative inner products
induced by the kernels $K_i,i=c,a,T$ will be disscussed in the next
section. For the moment note that there are plenty of them in each sector from the adiacent ones. The "massles" case in which the measure is $d\rho (p)=\theta (-p_0)\delta ^2 
(p^2 ) $ i.e. it is concentrated on the light cone deserves special attention. By putting together the non-negative inner products
$(.,.)_i ,i=c,a,T $ all zero vectors simply dissappear (see Section 3). We will
construct the natural unique supersymmetric positive definite scalar
product and obtain in the next section our rigorous Hilbert-Krein decomposition of the set of supersymmetric functions where the conditions (2.64) to (2.67) will play a central role.

\section{Hilbert-Krein Superspace}

In this section we present, on the basis of the results of Section 2, the generic Krein structure of
supersymmetries. Let $V$ be an inner product space with inner product
$<.,.>$ and $\omega $ an operator on $V$ with $\omega^2=1$ (do not
confuse this $\omega $ with the one in (2.67)). If
$(\phi,\psi)=<\phi,\omega \psi> ; \phi ,\psi \in V$ is a (positive
definite) scalar product on $V$ than we say that $V$ has a Krein
structure. By completing in the scalar product (.,.) we obtain an
associated Hilbert space structure (if $(.,.)$ has zero vectors we have
in addition to factorize them before completing). We obtain what we call 
a Hilbert-Krein space (or Hilbert-Krein structure). Hilbert-Krein
structures naturally appear in gauge theories (including the well
understood case of electrodynamics; see for instance the book \cite{Stro}).\\
Suppose the condition (2.48) on the measure $d\rho (p)$ is
satisfied ans, as always, $X$ and $Y$ are concentrated on its support. We decompose $X=X_1 +X_2 +X_3 $ where $X_1=X_c =P_c X ,X_2 =X_a
=P_a X, X_3=X_T =P_T X $. Then the simplest supersymmetric Hilbert-Krein structure which emerges from the considerations of the preceding section is given by
\begin{equation}
<X,Y>=\int d^8z_1d^8z_2\bar X^T (z_1)K_0 (z_1-z_2)Y(z_2)
\end{equation}
in the notations $X^T=(X_1 ,X_2 ,X_3 ), Y=\begin{pmatrix}Y_1 \\Y_2 \\Y_T \end{pmatrix}, K_0(z)=K_0
(z)I_3 $. Here  $I_3$ is the 3x3 identity matrix and $X^T $ is the
transpose of $X$. \\
Now let

\begin{equation}
(X,Y)=<X,\omega Y>
\end{equation}
with 

\begin{equation}\nonumber
\omega=\begin{pmatrix}1& 0 &0 \\0 &1& 0 \\0& 0 & -1 \end{pmatrix}
\end{equation}
Certainly $(.,.)$ is positive definite on the basis of results obtained
in Section 2. It is clear that although each inner product $(.,,)_i $
has zero vectors this will be no longer the case for (3.2).\\
Although very general the scalar product (3.4) is obstructed by the
(from the point of view of applications) unnatural restriction (2.48) of
the measure $d\rho $. It holds for the massive but fails for the massless case. Now the restrictions (2.62)-(2.67) on (test) supersymmetric functions come into play. Indeed, under these conditions
we can always decompose a supersymmetric function into its chiral, antichiral and transversal part and write down the indefinite as well as 
the definite scalar products (3.1) and (3.2). Note that in the massless case there is an overlap between chiral/antichiral and transversal sectors which consists of zero vectors and has to be factorized. From (2.11)-(2.13) follows that a function $X$ belongs to this overlap if

\begin{gather}\nonumber
X(z)=f(x)+\theta \varphi (x)+\bar \theta \bar \chi (x)\pm i\theta \sigma_l \bar \theta \partial_l f(x)
\end{gather}
with
\[\partial_l \varphi \sigma^l =\sigma^l \partial_l \bar \chi =0,\quad \square f=0 \] 
The restrictive condition on the measure was transfered to a restrictive
conditions on (test) functions, a procedure which is common for
rigorous quantum gauge fields (see for instance \cite{Stro}). In Section 2
we have seen that the content of the restrictive conditions on test
functions in supersymmetry might be of less extent as compared to
similar conditions in the usual case (remember the zero  divergence condition
which comes for free). The Hilbert-Krein structure on supersymmetric
functions subjected or not to the conditions (2.64)-(2.67) is the main result
of this paper.\\ 
We believe that it justified to call standard Hilbert-Krein
supersymmetric space the space of supersymmetric functions with indefinite and (positive) definite inner products given 
as above by

\begin{gather} \nonumber  
<X,Y>=\int \bar X^T K_0 Y \\ 
(X,Y)=<X,\omega Y>
\end{gather}
It is exactely the supersymmetric analog of the relativistic Hilbert
space used in quantum field 
theory in order to produce the Fock space of the free theory \cite{StreW}.
As a first application we mention here that the free chiral/antichiral
supersymmetric quantum field theory (i.e. the quantum field formally
generated by the free part of the Wess-Zumino Lagrangian) is
characterized by the posivive definite (at this stage only non-negative) 
two point function

\begin{gather}
\begin{pmatrix} \frac{1}{16}\bar D^2 D^2 & \frac{m}{4}\bar D^2 \\ \frac{m}{4}D^2 & \frac{1}{16}D^2 \bar D^2
\end{pmatrix}K_0
\end{gather}
where $d\rho (p)=\theta (-p_0)\delta (p^2+m^2)dp$ with $m>0$. 
The correspondence to the two-point functions of the chiral $\Phi$ and
antichiral $\bar \Phi $-quantum fields is indicated bellow 

\begin{gather}
\begin{pmatrix} \Phi \bar \Phi & \Phi \Phi \\ \bar \Phi \bar \Phi & \bar 
  \Phi \Phi \end{pmatrix}\sim \begin{pmatrix} \frac{1}{16}\bar D^2 D^2 & \frac{m}{4}\bar D^2 \\ \frac{m}{4}D^2 & \frac{1}{16}D^2 \bar D^2
\end{pmatrix}K_0
\end{gather}
The proof
of non-negativity of (3.4) is by computation. The factorisation of the
zero-vectors in (3.4) can be made explicit by imposing the equations of motion $\bar D^2\Phi =4m\Phi ,D^2 \Phi =4m\bar\Phi $ on the test functions \cite{C}. \\
The supersymmetric vacuum coincide with the function one and the
supersymmetric Fock space is symmetric (note that following our
reasoning all supersymmetric Fock spaces must be symmetric;
we expect antisymmetric Fock spaces for ghost fields).\\
A non-interacting quantum (free) system consisting of a chiral/antichiral and a
(massive) vector part is characterized by the positive definite operator 
in the standard Hilbert-Krein space (remember $T=-8\square P_T =D^\alpha 
\bar D^2 D_\alpha =\bar D_{\dot \alpha }D^2  D^{\dot \alpha} $) 

\begin{gather}
\begin{pmatrix} \frac{1}{16}\bar D^2 D^2 & \frac{m}{4}\bar D^2 & 0 \\
  \frac{m}{4}D^2 & \frac{1}{16}D^2 \bar D^2 & 0 \\ 0 & 0 & \frac{1}{8}T
\end{pmatrix}K_0
\end{gather}
The massless case is different; one has to take into account the conditions (2.64)-(2.67) \cite{C}. Oter applications include a supersymmetric K\"allen-Lehmann representation \cite{C}. \\

Before ending let us make two remarks. The first concerns the 
perspective of the present work. We succeeded to uncover the inherent
Hilbert-Krein structure of the $N=1$ superspace. It means that the
formal decomposition of supersymmetric functions into chiral, antichiral and transversal components, which was common tool from the first
days of superspace, was turned here into what we call the 
Hilbert-Krein structure of the $N=1$ superspace or the standard
supersymmetric Hilbert-Krein space. It shows that positivity 
(and as such unitarity) requires the substraction of the transversal
part instead of its addition as this might be suggested by the above
mentioned formal decomposition. Problems with the d'alembertian in the
denominator of the projections $P_i ,i=c,a,T $ have been disscussed. The 
natural way to avoid singularities is to impose some restrictions on the 
(test) functions. There are other applications in sight to which
we hope to come to (for some first modest steps see \cite{C}). \\
The second
remark is of technical nature. We worked in the frame of the van der
Waerden calculus using Weys spinors. This is very rewarding from the
point of view of computations in supersymmetry but is not totally
satisfactory from the rigorous point of view. Indeed the components of
the Weyl spinors as
coefficient functions for our supersymmetric (test) functions are
supposed to anticommute and this is unpleasant when tracing back the
supersymmetric integrals to usual $L^2 $- integrals. Of course this is
not a problem. A reformulation of the results using anticommuting Grassmann variables but commuting fermionic components is possible. The net results remain unchanged as it should be.

\end{document}